\documentstyle{mn}
\input epsf.tex

\def\mag{^{\rm m}}
\def\fm{\hbox{$.\!\!^{\rm m}$}}
\def\farcm{\hbox{$.\mkern-4mu^\prime$}}
\def\kms{\rm km\,s^{-1}}
\def\Ie{{< \hspace{-3pt} I \hspace{-3pt}>_{\rm e}}}
\def\Dn{{D_{\rm n}}}
\def\re{{r_{\rm e}}}
\def\mue{{{< \hspace{-3pt} \mu \hspace{-3pt}>}_{\rm e}}}
\def\mT{{m_{\rm T}}}
\def\Mgtwo{{ {\rm Mg}_2}}
\def\Fe{{< \hspace{-3pt} {\rm Fe} \hspace{-3pt}>}}
\def\Hb{{ {\rm H}{\beta}}}
\def\HbG{{ {\rm H}{\beta}_{\rm G}}}
\def\MV{{M_{\rm V}}}
\def\Ho50{{H_{\rm 0}=50~{\rm km\, s^{-1}\, Mpc^{-1}} }}
\def\AV{{A_{\rm V}}}
\def\AB{{A_{\rm B}}}
\def\Ar{{A_{\rm r}}}
\def\Teff{{T_{\rm eff}}}
\def\sigred{{\sigma _{\rm E(b-y)}}}
\def\sigmv{{\sigma _{\MV}}}

\title[The poor cluster of galaxies S639]
{The poor cluster of galaxies S639}
\author[I. J{\o}rgensen, H. J\o nch-S\o rensen]
{Inger J{\o}rgensen$^{1, }$\thanks{
E-mail: inger@roeskva.as.utexas.edu, helge@astro.ku.dk
\newline $\dagger$ Hubble Fellow. }$^{\,\dagger\, }$ and 
Helge J\o nch-S\o rensen$^{2, }$\mbox{\Huge $^{\star}$} \\
$^{1}$McDonald Observatory, The University of Texas at Austin, RLM 15.308, Austin, TX 78712, USA \\
$^{2}$Copenhagen University Observatory, Juliane Maries Vej 30, DK-2100 Copenhagen \O , Denmark }
\date{March 11, 1998, accepted for publication in MNRAS}

\begin{document}

\maketitle

\begin{abstract}
We have studied the poor southern cluster of galaxies S639.
Based on new Str\"{o}mgren photometry of stars in the direction of
the cluster we confirm that the galactic extinction affecting 
the cluster is large. We find the extinction in Johnson B to be
$\AB = 0.75\pm 0.03$.
We have obtained new photometry in Gunn $r$ for E and S0 galaxies
in the cluster. 
If the Fundamental Plane is used for determination of the relative
distance and the peculiar velocity of the cluster we find
a distance, in velocity units, of $(5706\pm 350)\kms$,
and a substantial peculiar velocity, $(839\pm 350)\kms$.
However, the colors and the absorption line indices of the E and 
S0 galaxies indicate that the stellar populations in these galaxies 
are different from those in similar galaxies in the two rich clusters 
Coma and HydraI.
This difference may severely affect the distance determination and the
derived peculiar velocity.
The data are consistent with a non-significant peculiar velocity 
for S639 and the galaxies in the cluster being on average 0.2 dex 
younger than similar galaxies in Coma and HydraI.
The results for S639 caution that some large peculiar velocities
may be spurious and caused by unusual stellar populations.

\end{abstract}


\section{Introduction}

The relation known as the Fundamental Plane (FP) may be used for 
determination of relative distances to E and S0 galaxies
(e.g., Dressler et al.\ 1987; J\o rgensen, Franx \& Kj\ae rgaard 1996, 
hereafter JFK96; Baggley 1996; Hudson et al.\ 1997).
The FP relates the effective radius, $\re$, the mean surface
brightness within this radius, $\Ie$ and the (central) velocity
dispersion $\sigma$, in a relation, which is linear in logarithmic
space (Djorgovski \& Davis 1987; Dressler et al.\ 1987).

The FP has a low scatter ($15-20$\% in $\re$) and is 
therefore a valuable tool for studies of peculiar velocities of 
galaxies and clusters (e.g., Baggley 1996; Hudson et al.\ 1997).
The use of the FP for determination of distances and
peculiar velocities relies on the 
assumption that the FP is universally valid. 
Several authors have investigated possible differences in the FP 
related to the cluster environment
(e.g., Burstein, Faber \& Dressler 1990; Lucey et al.\ 1991; 
de Carvalho \& Djorgovski 1992; JFK96; Baggley 1996).
Only de Carvalho \& Djorgovski find that the environment has 
significant effects on the FP. 
These authors find field galaxies to be brighter than cluster 
galaxies of similar effective radii and velocity dispersions.
de Carvalho \& Djorgovski also find field galaxies to be bluer and 
have weaker $\Mgtwo$ line indices than cluster galaxies with
similar velocity dispersions.
This is in general agreement with studies that show
that E and S0 galaxies in the outer
parts of clusters have weaker $\Mgtwo$ and $\Fe$ indices than
those in the central parts of clusters (Guzm\'{a}n et al.\ 1992; 
JFK96; J\o rgensen 1997).

In this paper we study the poor cluster of galaxies S639,
previously studied by JFK96.
The cluster identification is from Abell, Corwin \& Olowin (1989).
The cluster has a radial velocity in the Cosmic Microwave Background
(CMB) frame of $cz_{\rm CMB}=6545\kms$
and is located $\approx 28$$\degr$ from the direction to the
large mass-concentration known as the ``Great Attractor''
(Faber \& Burstein 1988).
The velocity dispersion of the cluster is $456_{-74}^{+83}\kms$ (JFK96).
Its richness is 14 measured as the number of galaxies with magnitudes
between $m_3$ and $m_3+2$ (Abell et al.\ 1989). $m_3$ is the magnitude
of the third ranked galaxy.
S639 has a smaller velocity dispersion and is poorer than clusters like
the Coma cluster and the HydraI cluster.
Coma and HydraI have velocity dispersions of $1010_{-44}^{+51}\kms$
and $608_{-39}^{+58}\kms$, respectively (Zabludoff, Huchra \& Geller 1990).
The richnesses given by Abell et al.\ (1989) is 106 for Coma and 39
for HydraI.

Using the FP for 10 E and S0 galaxies in S639 JFK96 found a large 
peculiar velocity of the cluster,
$v_{\rm pec}=(1295 \pm 359)\kms$ relative to the CMB frame.
Further, JFK96 found that the galaxies in the cluster follow a
$\Mgtwo$-$\sigma$ relation offset from the relation established
for their full sample of 11 clusters. The galaxies in S639 had on
average weaker $\Mgtwo$ indices, see also J\o rgensen (1997).
JFK96 tried to correct the derived peculiar velocity of the cluster
for the offset in the $\Mgtwo$ indices by including a $\Mgtwo$ term
in the FP. The result was $v_{\rm pec}=(879\pm 392)\kms$. 
However, the coefficient for the $\Mgtwo$ term is not well determined,
cf.\ JFK96.

S639 is located at low galactic latitude, 
($l$,$b$) = ($280\degr$,$11\degr$). Thus, the galactic extinction is
large and uncertainties in the adopted value may severely affect
the precision of the derived distance and peculiar velocity
for the cluster.

The main issue discussed in this paper is whether the large peculiar 
velocity of S639 found by JFK96 is real or the result was caused 
either by incorrect correction for the (large) galactic extinction,
by selection effects, or by unusual stellar populations.
In order to reach conclusions about these issues we have obtained
additional photometry of galaxies in S639, giving a sample of 
21 E and S0 galaxies with available photometric and spectroscopic
parameters. 
We have also obtained Str\"{o}mgren $uvby$-$\beta$
photometry for stars in the direction of the cluster.
This photometry is used to determine the galactic extinction
affecting the cluster. 

The sample selection for the E and S0 galaxies and the available data 
are briefly described in Sect.\ 2.
The determination of the galactic extinction is covered in Sect.\ 3.
In Sect.\ 4 the FP is discussed and used for determining the
distance to the cluster.
The importance of the stellar populations is investigated in Sect.\ 5.
The conclusions are summarized in Sect.\ 6.

The relations between the parameters for the galaxies established
in this paper are determined by minimization of the sum of the 
absolute residuals perpendicular to the relations.
This fitting technique has the advantage that it is rather insensitive
to a few outliers, and that it treats the coordinates in a 
symmetric way. The uncertainties of the coefficients are derived
by a bootstrap method.
See also JFK96 for a discussion of this fitting technique.

\section{Sample selection and data}

The sample of E and S0 galaxies in the poor cluster S639
used in the analysis by JFK96 was selected based on sky survey images. 
This sample was by no means complete. We have obtained additional
photometry in Gunn $r$ of galaxies within the central
$25' \times 35'$ of the cluster. 
We aimed at constructing a magnitude limited sample of E and S0
galaxies from the photometry from J\o rgensen, Franx \& Kj\ae rgaard 
(1995a, hereafter JFK95a), our new photometry in Gunn $r$ 
and the spectroscopic data from 
J\o rgensen, Franx \& Kj\ae rgaard (1995b, hereafter JFK95b)
and J\o rgensen (1997).
We classified the galaxies based on our CCD images. For galaxies
outside the central area the classifications from JFK95a were
adopted.
In order to further check our classifications we subtracted models
of the best fitting elliptical isophotes from the images of the
galaxies. None of the galaxies in the sample show residual spiral
arms after the model subtraction.
Thus, we find it unlikely that our sample is con\-ta\-minated by
early-type spirals.
The accidential inclusion of a few early-type spirals in the fainter
part of the sample, for which the residual spiral arms may be difficult
to detect, will not affect our results significantly, since none of
the results depends critically on the faintest third of the sample.

The final sample contains 21 E and S0
galaxies for which both photometric and spectroscopic parameters are
available.
Within the central $25' \times 35'$ the sample is 90\% complete to a 
total magnitude of $15\fm 5$ in Gunn $r$ (magnitudes corrected for 
galactic extinction, k-corrected and corrected for cosmological 
dimming).
Further, we have data for two fainter galaxies, and two galaxies
outside the central area. 

\subsection{Available data for the galaxies in S639}

The new photometry of S639 was obtained with the 
Danish 1.5-m telescope, LaSilla, March 22--28, 1997. 
We used the DFOSC (Danish Faint Object Spectrograph 
and Camera) equipped with a 2k$\times$2k thinned Loral CCD.
The data were reduced with standard methods. 
Two-dimensional surface photometry of the galaxies and the effective 
parameters were derived in the same way as done by JFK95a.
The magnitudes of the new data were standard calibrated by means of 
zero point offsets relative to the photometry from JFK95a. 
The comparison of the aperture magnitudes
with the data from JFK95a has an rms scatter of $0\fm 01$.
More details of the data reduction are given in Appendix A.

JFK95a derived the $(g-r)$ colors for 10 of the E and S0 galaxies in 
S639. The colors were measured within the effective radii in Gunn $r$.
We have transformed the $(g-r)$ colors to $(B-r)$ using the relation
$(B-r) = 0.88 (g-r)+0.73$. We established the relation from
color data for a total of 41 E and S0 galaxies in the
clusters Abell 194, DC2345-28, and HydraI (JFK95a;
Milvang-Jensen \& J\o rgensen 1998).
The rms scatter of the relation is 0.025.

We use spectroscopic data from JFK95b and J\o rgensen (1997). 
Velocity dispersions are available for 21 E and S0 galaxies.
The absorption line indices $\Mgtwo$ and $\Fe$ have been measured
for 20 of those galaxies, and four galaxies have measured $\HbG$.
The $\Mgtwo$ and $\Fe$ line indices are on the Lick/IDS system
(named after the Lick Image Dissector Scanner;
Faber et al.\ 1985; Worthey et al.\ 1994). 
The $\HbG$ index is related to the Lick/IDS $\Hb$ index as 
$\HbG = 0.866 \Hb + 0.485$ (J\o rgensen 1997).
All the spectroscopic parameters are centrally measured values 
corrected to a circular aperture with a diameter of 
1.19 h$^{-1}$\,kpc (JFK95b; J\o rgensen 1997),
$H_{\rm 0} =100\,{\rm h}\,{\rm km\,s^{-1}\,Mpc^{-1}}$.
The line indices are corrected for the effect of the velocity
dispersion.

The photometric and spectroscopic parameters for the galaxies
in S639 are given in the Appendix A.

\subsection{Reference samples in other nearby clusters}

We use the Coma cluster and the HydraI cluster as reference clusters. 
The photometric data for the Coma cluster are from JFK95a.
The spectroscopic data are from the literature (Davies et al.\ 1987;
Dressler 1987; Lucey et al.\ 1991; Guzm\'{a}n et al.\ 1992) and have
been calibrated to a consistent system by JFK95b.
We also use new spectroscopic data for additional galaxies in the
Coma cluster (J\o rgensen 1998).
The data for the HydraI cluster are from
JFK95a, JFK95b and Milvang-Jensen \& J\o rgensen (1998).

The sample of E and S0 galaxies in Coma covers the central
$64' \times 70'$ of the cluster and is 93\% complete
to a magnitude limit of $15\fm 05$ in Gunn $r$.
This sample contains 116 galaxies.
The sample of E and S0 galaxies in HydraI covers the central
$46' \times 76'$ of the cluster and is 80\% complete to a 
limit of $14\fm 5$ in Gunn $r$. This sample contains 45 galaxies.
Both samples include a few galaxies below their magnitude limit.
The line indices $\Fe$ and $\HbG$ are available for sub-samples
of the samples in Coma and HydraI.

There are three reasons to use these large samples of galaxies in Coma
and HydraI as the reference samples, rather than making use of
all the clusters studied by JFK96.
First, our S639 sample is magnitude limited and we therefore
expect more accurate results by comparing this sample to other
magnitude limited sampes, rather than using the incomplete samples
from JFK96.
Second, the JFK96 sample contains two clusters with smaller velocity
dispersions than S639 and comparable offsets in the $\Mgtwo$ index
relative to the Coma cluster. Thus, including these clusters in
the reference sample would make that sample inhomogeneous.
Third, our goal is to discuss the reality of the large peculiar
velocity of S639 found by JFK96. JFK96 derived peculiar velocities
under the assumption that Coma is at rest relative to the CMB frame.
In this paper we make the same assumption in order to facilitate
the comparison with the results from JFK96.

\section{The galactic extinction in the direction of S639}

The galactic extinction affecting observations of extragalactic
objects is traditionally determined from the HI-mapping done by 
Burstein \& Heiles (1982).
From this mapping we obtain a galactic extinction in the direction
of S639 of $\AB = 0.80$.
However, in some areas especially near the galactic plane the
extinction has been found to be different from estimates based
on the Burstein \& Heiles maps (e.g., Burstein et al.\ 1987;
J\o nch-S\o rensen 1994b).
In the following we therefore use two other methods to estimate
the galactic extinction in the direction of S639: Str\"{o}mgren
photometry of stars in the direction of the cluster, and
the color-$\Mgtwo$ relation for galaxies in the cluster.

\begin{figure}
\epsfxsize=8.8cm
\epsfbox{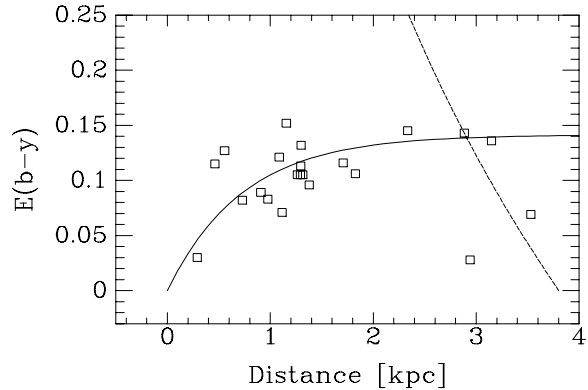}

\caption[]{
The galactic reddening $E(b-y)$ in the direction of S639.
The 22 stars with $2.58 \le \beta \le 2.72$ and
computable $E(b-y)$ and $\MV$ are shown. 
The solid curve is a model of the reddening caused by the diffuse 
interstellar gas. The dashed curve marks the maximum observable 
reddening at a limiting magnitude of $V=17\mag$ and $\MV = 4\fm 1$,
corresponding to the limiting magnitude and mean $\MV$ of 
the observed sample of stars.  \label{fig-Ebydist}
}
\end{figure}

\subsection{The Str\"{o}mgren photometry}

S639 was observed as part of a program to obtain Str\"{o}mgren 
$uvby$-$\beta$ photometry of stars in the directions of several galaxy 
clusters at low galactic latitude.
The observations were obtained with the Danish 1.5-m telescope,
La Silla, March 13--29, 1996.
The telescope was equipped with the DFOSC.
The reduction and calibration of the photometry are described
in detail in Appendix B.
Here we concentrate on the determination of the galactic
extinction in the direction of S639.

Intrinsic colors can be estimated in the $uvby$-$\beta$ system for
individual stars of spectral types ranging from B- to G-type.
This has been used extensively to map the reddening in the solar 
neighborhood (Hilditch, Hill \& Barnes 1983; Franco 1989).
Recently deep CCD-photometry surveys have made it possible to trace 
reddening to large distances, hence all the way through the 
galactic dust disk using individual stars 
(J\o nch-S\o rensen 1994b; J\o nch-S\o rensen \& Knude 1994).
Nissen (1994) reviewed the calibrations and the accuracy of the 
derived parameters. He showed that the rms scatter of the
residuals in the calibrations sets the limit of the precision with
which $E(b-y)$ can be derived to 0.009. This corresponds to
an uncertainty of 0.012 in $E(B-V)$ and 0.050 in $\AB$.
We use $E(B-V) = 1.35E(b-y)$, $\AV = 4.2E(b-y)$, and $\AB = 5.57E(b-y)$
to convert between reddening and galactic extinction.

In Fig.\ \ref{fig-Ebydist} distance versus $E(b-y)$ is shown for the 
stars in the direction of S639.
The typical uncertainty on $E(b-y)$ is $\approx 0.04$.
The relative distance uncertainty is $\la$30\% for $V\la 16\mag$ and
increasing rapidly for fainter stars, see Appendix B.
The mean reddening $E(b-y)$ for stars at distances larger than 750\,pc 
is 0.106 with an rms scatter of 0.031.
The solid curve in Fig.\ \ref{fig-Ebydist} represents the predicted 
$E(b-y)$-distance relation for a model from J\o nch-S\o rensen (1994b) 
with a vertical scale height of the (diffuse) absorbing material of 
140\,pc and a local normalization of 0.42\,atoms\,cm$^{-3}$.
For other parameters of the model see J\o nch-S\o rensen (1994b). 
The model indicates a total reddening of $E(b-y) = 0.14$, 
equivalent to $\AB = 0.78$, for objects situated outside the disk. 
From this we find the visual absorption acquired through half the 
diffuse disk in this direction of the Galaxy to be $\AV \sin b = 0.11$
($b$ in this expression is the galactic latitude). 
This value can be compared with the results for the
seven fields surveyed by J\o nch-S\o rensen (1994b) and 
J\o nch-S\o rensen \& Knude (1994).
The result for the S639 direction is comparable with $\AV \sin b =0.13$
found for the low latitude field ($l$,$b$)=(262$\degr$,+4$\degr$),
but lower than the 0.21 found for ($l$,$b$)=(270$\degr$,+35$\degr$).

\begin{figure}
\epsfxsize=16.5cm
\hspace*{-0.1cm}\vspace*{-0.6cm}
\epsfbox{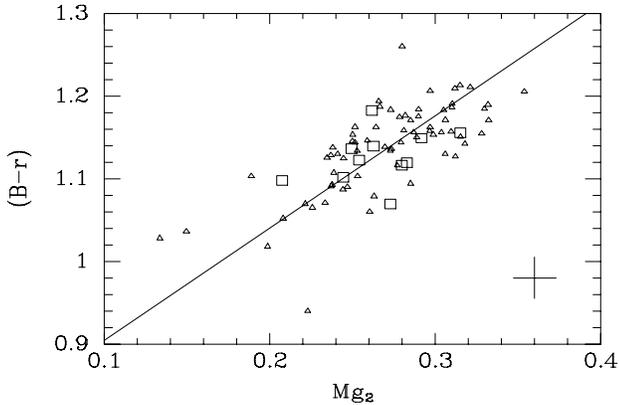}

\vspace*{-4.8cm}

\caption[]{
The $(B-r)$-$\Mgtwo$ relation.
Large boxes -- S639; small triangles -- Coma and HydraI.
Typical error bars are given on the figure.
The solid line is the relation derived from Coma and HydraI,
$(B-r) = (1.36\pm 0.22)\Mgtwo + 0.77$.
The rms scatter in $(B-r)$ for the galaxies in S639 is of 0.04.
The galactic extinction for S639 was assumed to be $\AB = 0.72$.
This gives consistency with the $(B-r)$-$\Mgtwo$ relation for 
the Coma and HydraI galaxies.
\label{fig-BrMg}
}
\end{figure}

The reddening caused by the diffuse interstellar medium may be 
understood as contributions from both diffuse interstellar clouds and 
an intercloud medium (e.g., Knude 1979b).
The typical reddening due to a diffuse cloud is $E(b-y) \approx 0.03$
and the average number of clouds intercepted is 4.3 per kpc 
(Knude 1979a, 1981).
The intercloud medium contributes with only $E(b-y) \approx 0.001$ per
100 pc (Knude 1979b).
In total this yields a reddening  of $E(b-y) = 0.14$\,kpc$^{-1}$ 
along a line of sight in the disk. 
In the direction of S639 a distance of 1 kpc corresponds to a 
height above the plane of 190\,pc, or about 1.5 times the 
scale height of the diffuse medium. 
Thus, based on the results from local data we
would expect the reddening towards S639 due to the diffuse interstellar
medium to be close to $E(b-y)=0.14$, as indicated by Fig.\ 1.  
Two distant stars in the observed field show low reddening. 
This may be a result of the `stochastic' nature of the distribution of 
the clouds, or simply due to the rather large measurement 
errors on $E(b-y)$ for the individual stars. 
The two stars have reddenings of approximately 0.03 and 0.07, and may
represent lines of sight that intercept one and two clouds,
respectively.

The sampling and accuracy of the present results do not allow
for a specific mapping of clouds or tracing of reddening variations 
across the field. 
The best estimate of the typical reddening affecting the
observations of the galaxies in S639 is $E(b-y)= 0.14$, 
corresponding to $E(B-V) = 0.19$, $\AV = 0.59$ and $\AB = 0.78$.

\begin{figure*}
\epsfxsize=16.5cm
\hspace*{0.8cm}
\epsfbox{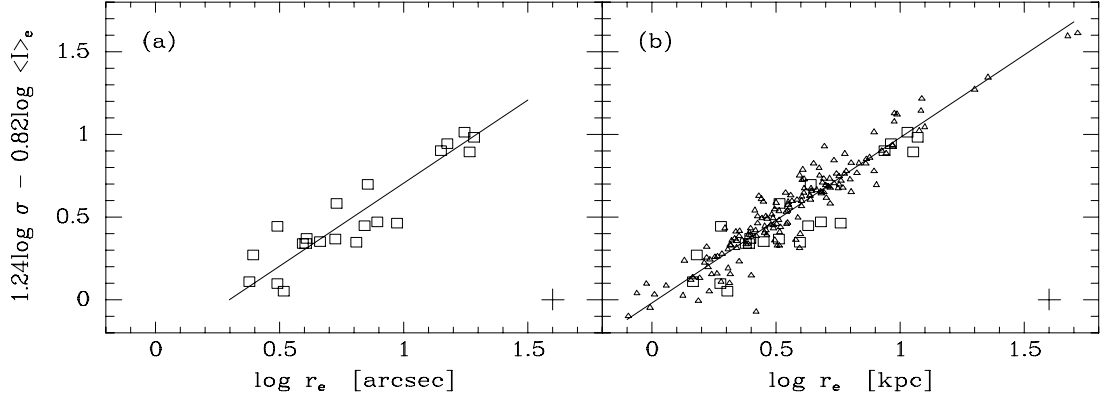}

\vspace*{-4.8cm}

\caption[]{
The FP for S639 seen edge-on. The photometry is in Gunn $r$,
$\AB=0.75$ was used.  (a) The effective radii in arc seconds. 
The solid line is the FP with the zero point derived from the S639
galaxies. 
(b) The effective radii in kpc, under the assumption of zero peculiar
velocity.  Large boxes -- S639; small triangles -- Coma and HydraI.
The solid line is the FP with the zero point defined by the
E and S0 galaxies in Coma.
Typical error bars are given on the panels.
\label{fig-fp}
}
\end{figure*}

\subsection{The $(B-r)$\,-\,$\Mgtwo$ relation}

E and S0 galaxies in nearby clusters are known to follow a relation 
between the color and the $\Mgtwo$ index (Burstein et al.\ 1988;
Bender, Burstein \& Faber 1993).
If we assume that the relation is universal, then it may be used as 
an independent way of estimating the galactic extinction. 
The assumption means that we expect it to be a universal
property of these galaxies that the global stellar populations 
match the central stellar populations (cf., Bender et al.\ 1993),
since the colors are measured within large apertures (the effective
radii) while the $\Mgtwo$ indices are measured within much smaller
apertures.
The color-$\Mgtwo$ relation was used by Burstein et al.\ (1987) to 
estimate the galactic extinction affecting galaxies at some directions
close to the galactic plane.
The relation has also been used recently by Schlegel, Finkbeiner \& Davis
(1998) to calibrate the DIRBE/IRAS dust maps to galactic reddening
(DIRBE: Diffuse Infra Red Background Experiment on board the COBE 
satellite; IRAS: Infrared Astronomy Satellite).
Burstein et al.\ (1988) found the scatter in the color-$\Mgtwo$
relation to be compatible with the measurement errors. Therefore,
we do not expect the stellar population differences discussed in
Sect.\ 5 to have significant effect on the galactic extinction derived
from the color-$\Mgtwo$ relation. 

We assume that E and S0 galaxies in S639 follow the same
$(B-r)$-$\Mgtwo$ relation as E and S0 galaxies in Coma and HydraI.
The $(B-r)$-$\Mgtwo$ relation for Coma and HydraI is shown in 
Fig.\ \ref{fig-BrMg}.
The data for the E and S0 galaxies in S639 are overplotted.
With a galactic extinction of $\AB = 0.72$ for S639
the data for the cluster are consistent with the Coma and HydraI
$(B-r)$-$\Mgtwo$ relation. 
The rms scatter the $(B-r)$-$\Mgtwo$ relation for S639 is 0.04.
This gives an uncertainty on 
$\AB$ of $\pm 0.03$, equivalent to an uncertainty on the galactic
extinction in Gunn $r$, $\Ar$, of $\pm 0.02$.

In the following we use the average of $\AB$ derived from
the Str\"{o}mgren photometry and the $\AB$ determination
based on the $(B-r)$-$\Mgtwo$ relation.
We adopt $\AB =0.75\pm 0.03$ ($E(B-V)=0.18\pm 0.01$) for 
all the galaxies in S639,
where the uncertainty is half the difference between our two
determinations. The corresponding galactic extinction in 
Gunn $r$ is $\Ar = 0.47\pm 0.02$.
We note that our determination of $E(B-V)$ for S639 agrees with
the dust maps from Schlegel et al.\ (1998). These maps imply a
mean value of $E(B-V)=0.152\pm 0.024$ within the central 
$25'\times 35'$ of the cluster.

\section{The Fundamental Plane for S639}

In Fig.\ \ref{fig-fp}a we show the FP for S639 edge-on. 
The effective radii are in arcsec and the mean surface brightnesses 
are given as $\log \Ie = -0.4(\mue - 26.4)$. 
$\Ie$ has units of ${\rm L}_{\odot}\, {\rm pc^{-2}}$.
The solid line on the figure is the FP for nearby clusters 
found by JFK96,
$\log \re = (1.24\pm 0.07) \log \sigma - (0.82\pm 0.02) \log \Ie + \gamma$.
The zero point is defined by the present sample of galaxies in S639.  
The rms scatter for S639 is 0.122 in $\log \re$.

If we fit the FP to the S639 data we find
\begin{equation}
\begin{array}{lcl}
\log \re & = & \hspace*{7pt}1.32 \log \sigma -0.78 \log \Ie + 0.007 \\
         &   & \pm 0.30 ~~~~~~\hspace*{2pt} \pm 0.09
\end{array}
\label{eq-fps639}
\end{equation}
with an rms scatter 0.136 in $\log \re$.
Thus, the FP for S639 is not significantly different from the 
FP found for nearby clusters (JFK96), or the FP fitted to the Coma 
and HydraI samples (J\o rgensen 1998;
Milvang-Jensen, J\o rgensen \& J\o nch-S\o rensen 1998).
In the following we use the FP established by JFK96.

Fig.\ \ref{fig-fp}b shows the FP edge-on with the effective radii 
given in kpc ($\Ho50$). The data for S639 have been overplotted upon 
the data for the clusters Coma and HydraI. 
We have assumed that the peculiar velocity of S639 is zero.
The solid line on this figure is the FP with
the zero point defined from Coma. The offset between Coma and S639 is 
$0.100\pm 0.027$ in $\log \re$. This may be interpreted as due to
a non-zero peculiar velocity of S639.
We assume that the Coma cluster is at rest relative to the
CMB frame and has $cz_{\rm CMB} = 7200 \kms$ (JFK96).
We then find the distance of S639, expressed as the expected 
radial velocity, to be $(5706 \pm 350)\kms$.
This implies a peculiar velocity of $(839\pm 350)\kms$.
If we include a correction for the cluster's
offset in the $\Mgtwo$-$\sigma$ relation, see Sect.\ 5,
as also tried by JFK96, we find a distance of $(5989\pm 378)\kms$ and
$v_{\rm pec}=(556\pm 378)\kms$.
The results were not corrected for Malmquist bias. 
Based on the rms scatter of the FP we expect the Malmquist bias
from a homogeneous density field to be $\approx 1.3$\%, 
cf.\ Lynden-Bell et al.\ (1988).

For both methods the peculiar velocities of S639 found 
here are smaller than those found by JFK96. 
The differences are partly due to the larger sample of
galaxies included in the present study, and partly due to the 
slightly lower value assumed of the galactic extinction.
Our result confirms, however, that if the FP without any $\Mgtwo$
term is used for distance determination for S639 then a 
large positive peculiar velocity is found.

The rms scatter of the FP for S639 is slightly larger than found for the
Coma and the HydraI clusters (JFK96; J\o rgensen 1998; 
Milvang-Jensen et al.\ 1998).
Part of this may be caused by variation of the galactic extinction
over the field of S639. Our Str\"{o}mgren photometry for the
stars in the field is not accurate enough to investigate this. 
However, for other fields within 30 degrees of the galactic plane
values of the rms scatter in $E(b-y)$ of 0.035-0.10 have been
found (J\o nch-S\o rensen 1994b; J\o nch-S\o rensen \& Knude 1994).
If the field of S639 has similar variations in $E(b-y)$ it 
would result in a contribution
of $\approx 0.04$ to the rms scatter of the FP (measured in the 
direction of $\log \re$). This explains only a small part of the
scatter in the FP for S639, but it does bring the unexplained part
of the scatter, $\approx 0.10$ in $\log \re$, in better agreement
with recent results for Coma and HydraI (J\o rgensen 1998; 
Milvang-Jensen et al.\ 1998).

\begin{table}
\caption[]{Model predictions from Vazdekis et al.\ (1996)
\label{tab-model} }
\begin{tabular}{l@{\,$\approx$\ }l} \hline
$\Mgtwo$   & \hspace*{8pt}0.12\,log\,age + 0.19[M/H] + 0.14 \\
$\log \Fe$ & \hspace*{8pt}0.12\,log\,age + 0.25[M/H] + 0.34 \\
$\log \HbG$ & -- 0.27\,log\,age -- 0.135[M/H] + 0.51 \\ 
$\log M/L_r$ & \hspace*{8pt}0.63\,log\,age + 0.26[M/H] -- 0.16 \\ 
$(B-r)$ & \hspace*{8pt}0.36\,log\,age + 0.36[M/H] + 0.83 \\ \hline
\end{tabular}

Note -- [M/H]$\equiv \log Z/Z_{\odot}$ is the total metallicity 
relative to solar.
\end{table}

\begin{table}
\caption[]{Systematic offsets for S639 galaxies \label{tab-offset} }

\begin{tabular}{lccc}
Parameter    & Offset & $\Delta$ log\,age & $\Delta$[M/H] \\ \hline 
$\log M/L_r$ & $-0.12\pm 0.03$   & $-0.19\pm 0.05$ & $-0.46\pm 0.12$ \\
$(B-r)$      & $-0.045\pm 0.010$ & $-0.13\pm 0.03$ & $-0.13\pm 0.03$ \\
$\Mgtwo$     & $-0.021\pm 0.006$ & $-0.18\pm 0.05$ & $-0.11\pm 0.03$ \\
$\log \Fe$   & $-0.034\pm 0.019$ & $-0.28\pm 0.06$ & $-0.14\pm 0.08$ \\
$\log \HbG$  & $~~0.078\pm 0.015$  & $-0.29\pm 0.06$ & $-0.58\pm 0.11$ \\ \hline
\end{tabular}

Notes -- offsets calculated as the mean difference between the 
parameters for galaxies in S639 and the established relations.

\end{table}

\section{The stellar populations of the galaxies}

In this section we investigate whether the offset between the FP for
S639 and the FP for Coma and HydraI may be due to differences in 
the stellar populations in the galaxies.

\begin{figure}
\epsfxsize=16.5cm
\vspace*{-1.5cm}
\hspace*{-0.3cm}
\epsfbox{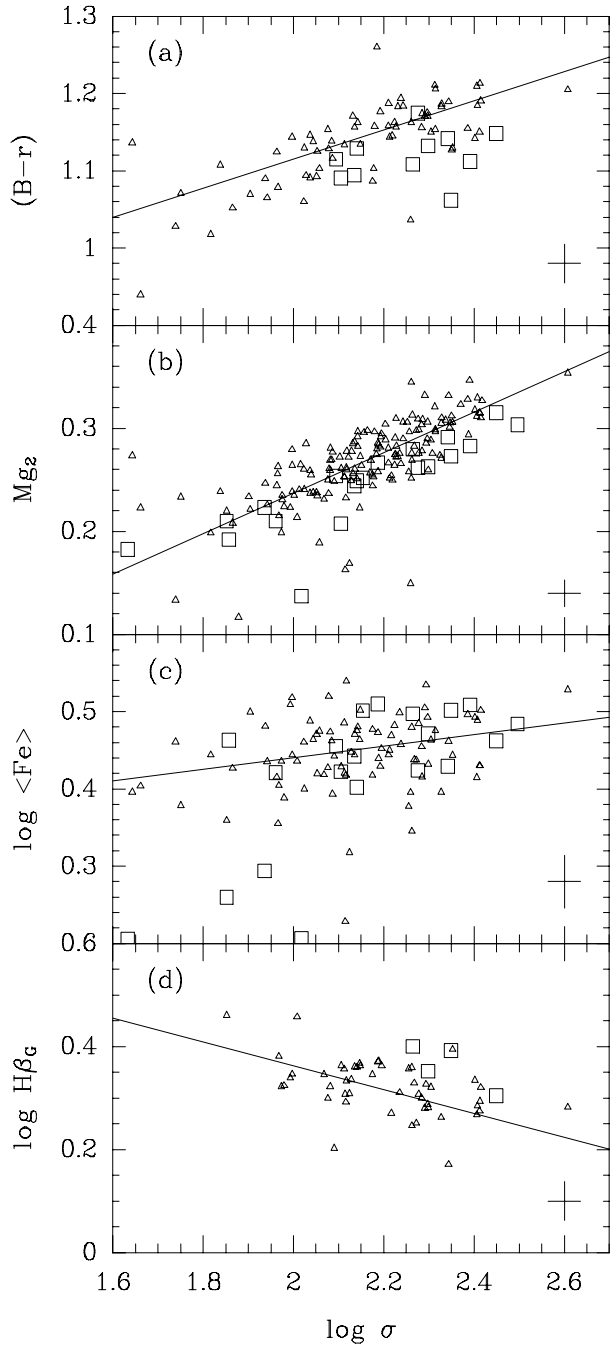}

\vspace*{-2.1cm}
\caption[]{
The color $(B-r)$ and the line indices $\Mgtwo$, $\Fe$ and
$\HbG$ versus the velocity dispersion.
Large boxes -- S639; small triangles -- Coma and HydraI.
Typical error bars are given on the panels.
The relation on panel (a) has been established from the 
Coma and HydraI samples. 
We find \mbox{$(B-r)$} $=(0.188\pm 0.044)\log \sigma - 0.740$ 
with an rms scatter of 0.037.
The relations on panels (b), (c) and (d) are adopted from J\o rgensen
(1997).
The galaxies in S639 are systematically bluer, have smaller 
$\Mgtwo$ indices, and larger $\HbG$ indices than the galaxies in 
Coma and HydraI.
A few of the galaxies in S639 also have substantially smaller
$\Fe$ indices than the bulk of the galaxies in Coma and HydraI.
\label{fig-MgFesigma}
}
\end{figure}

E and S0 galaxies show a  number of relations between the velocity
dispersions, the absorption line indices and the colors.
Differences in the stellar populations of galaxies in different clusters
are expected to show up as differences in these relations.
We establish the differences as zero point offsets for the S639
sample relative to the relations for the Coma and HydraI samples.
In order to interpret the offsets in terms of differences in 
the mean age and/or metallicity of the stellar populations we use
the single stellar population models from Vazdekis 
et al.\ (1996). Table \ref{tab-model} gives the expected changes
in the line indices, the M/L ratio and the color due to
changes in the age and/or metallicity.
These relations are valid for models with ages larger than 5\,Gyr
and a bi-modal initial mass function (IMF) with a high mass slope 
identical to a Salpeter (1955) IMF ($\mu$=1.35).

The relations between the velocity dispersions and the indices
$\Mgtwo$, $\Fe$ and $\HbG$ as well as between the velocity dispersion
and the color $(B-r)$ are shown in Fig.\ \ref{fig-MgFesigma}.
The \mbox{$(B-r)$-}$\sigma$ relation on Fig.\ \ref{fig-MgFesigma}a was
established from the Coma and HydraI galaxies, while the relations 
between the line indices and the velocity dispersion are adopted from 
J\o rgensen (1997).
The galaxies in Coma and HydraI follow these relations.

The galaxies in S639 are offset relative to the $(B-r)$-$\sigma$ 
relation, the $\Mgtwo$-$\sigma$ relation and the $\HbG$-$\sigma$
relation. The galaxies are on average bluer, have smaller 
$\Mgtwo$ indices and larger $\HbG$ indices than the galaxies in 
Coma and HydraI.
Table \ref{tab-offset} summarizes the offsets for all the relations.
The table also gives the offset for the FP in terms of an offset in
the M/L ratio, assuming that the peculiar velocity of S639 is zero.

We use the stellar population models to estimate
the change in age or metallicity that would create the 
measured offsets in the various parameters, see Table \ref{tab-offset}.
We assume that the offsets are either due to age variations, 
only, or metallicity variations only.
The offsets in $\Mgtwo$, $\HbG$, $(B-r)$ and the M/L ratio are
all consistent with a difference between the mean ages of the
galaxies in S639 and the galaxies in Coma and HydraI of --0.2 dex.
Though the $\Fe$ indices are consistent with this it is obvious
from Fig.\ \ref{fig-MgFesigma}c that the offset in $\log \Fe$ is
caused by four galaxies with very small $\Fe$ indices.
For the rest of the galaxies 
there is no significant offset in
$\Fe$, while these galaxies show the same offsets in $\Mgtwo$ and
$\log M/L_r$ as the full sample.
A possible reason may be that the galaxies in S639 have a lower
abundance ratio [Mg/Fe] than the bulk of E and S0 galaxies in
Coma and HydraI.
It requires better measurements of $\Fe$ to further investigate 
this potential problem with our interpretation. 

If we interpret the offsets as due to a systematic offset in the
metallicity, while there is no difference in the mean ages,
we find a mean difference of --0.12 dex, based on the offsets
in $(B-r)$ and $\Mgtwo$. 
However, a difference in the metallicity cannot fully explain the 
offset in the M/L ratio. In this case S639
may have a non-zero peculiar velocity of $\approx 490 \kms$.
We also note that the offset in $\HbG$ cannot be fully explained
by a metallicity difference.

\section{Conclusions}

The galactic extinction in the direction of the poor cluster of 
galaxies S639 has been determined from Str\"{o}mgren $uvby$-$\beta$
photometry for stars in the direction of the cluster.
Further, we have tested the consistency of the derived galactic
extinction by using the $(B-r)$-$\Mgtwo$ relation for E and S0
galaxies in the cluster. Our best estimate of the galactic extinction
in the direction of the cluster is $\AB = 0.75\pm 0.03$.

The FP for S639 has been established based on a sample of 21 E and
S0 galaxies. The coefficients for the FP for this cluster are 
not significantly different from those of the FP for the Coma and 
the HydraI clusters, and are also in agreement with previous results 
for other nearby clusters (e.g., JFK96).
Under the assumption that the FP (coefficients and zero point) 
is universal we find a distance, in velocity units,
to S639 of $(5706\pm 350) \kms$.
This implies a peculiar velocity for the cluster of
$(839\pm 350)\kms$.

The E and S0 galaxies in S639 have significantly smaller $\Mgtwo$
indices, larger $\HbG$ indices and are bluer than E and S0 galaxies 
of similar velocity dispersions in Coma and HydraI.
The offset in the FP for S639 relative to Coma and HydraI may be
due to a difference in the stellar populations, rather than a
large peculiar velocity for S639.
The data are consistent with a zero peculiar velocity and
mean ages of the S639 galaxies 0.2 dex younger than the
mean ages of similar galaxies in Coma and HydraI.
Alternatively, the $\Mgtwo$ indices and the $(B-r)$ colors 
are consistent with a metallicity difference of 0.1 dex, with
galaxies in S639 having a lower metallicity than those in 
Coma and HydraI.
In this case the peculiar velocity of S639 is $\approx 490\kms$.
However, this interpretation is not consistent with the strong 
$\HbG$ indices measured for the four galaxies, for which we 
have measurements of this index.

We conclude that the peculiar velocity of S639 is most likely 
overestimated if the FP is used as a distance determinator for this 
cluster.
Even though many studies have shown that the FP (and the $\Dn$-$\sigma$
relation which is a projection of the FP) to a large degree is 
universally valid 
(e.g., Burstein, Faber \& Dressler 1990; JFK96; Baggley 1996),
our results for S639 caution that there may be exceptions
(see also Gregg 1992).
When distance determinations are attempted the best approach 
will be to obtain
colors and line indices together with the other required data.
This will give the possibility of identifying clusters (and galaxies),
which deviate strongly from the mean relations between the 
various global parameters. These clusters can also be expected to
deviate from the FP otherwise valid for the bulk of the E and S0 
galaxies.
One may attempt to include a $\Mgtwo$ term in the FP in order to
correct for the effects caused by differences in the stellar populations.
However, the coefficient for such a term is not well determined, cf.\ JFK96,
and the peculiar velocities derived with this method may not be accurate
enough for investigations of large-scale flows.

\vspace{0.5cm}
Acknowledgements:
Lars Freyhammer is thanked for obtaining part of the observations
used for this research.
The Danish Board for Astronomical Research and the European Southern 
Observatory are acknowledged for assigning observing time
for this project and for financial support.
Support for this work was provided by NASA through grant
number HF-01073.01.94A to IJ from the Space Telescope Science Institute,
which is operated by the Association of Universities for Research
in Astronomy, Inc., under NASA contract NAS5-26555.
HJS acknowledges financial support from the Carlsberg Foundation,
Denmark.

%
 
\appendix
 
\section{Photometry for the galaxies}
 
Observations were obtained for five CCD fields, covering
the central $25'\times 35'$ of S639.
The exposure time for each field was two times 5 minutes in Gunn $r$.
The CCD images were reduced in a standard way. 
This includes correction for bias, and flat field correction with 
dome flat fields
with an additional correction for the low frequency variation 
derived from flat fields obtained at twi-light.
The pixel-to-pixel accuracy of the flat field correction is
0.6\%, while the accuracy of the correction for the 
low frequency variation is better than 1\%.

The two images of each field were registered and added before
determination of the photometric parameters of the galaxies.
The determination of the photometric parameters were done in the
same way as in JFK95a.
Two-dimensional surface photometry for the galaxies were derived
with the program package {\sc GALPHOT} (Franx, Illingworth \& Heckman
1989; J\o rgensen, Franx \& Kj\ae rgaard 1992).
This package fits a full 2-dimensional model to the image of a galaxy.
The results are radial profiles of the surface brightness, the 
ellipticity, the position angle and the deviations of the isophotes
from ellipses, as well as the growth curves of the galaxies.

\begin{figure}
\epsfxsize=8.8cm
\epsfbox{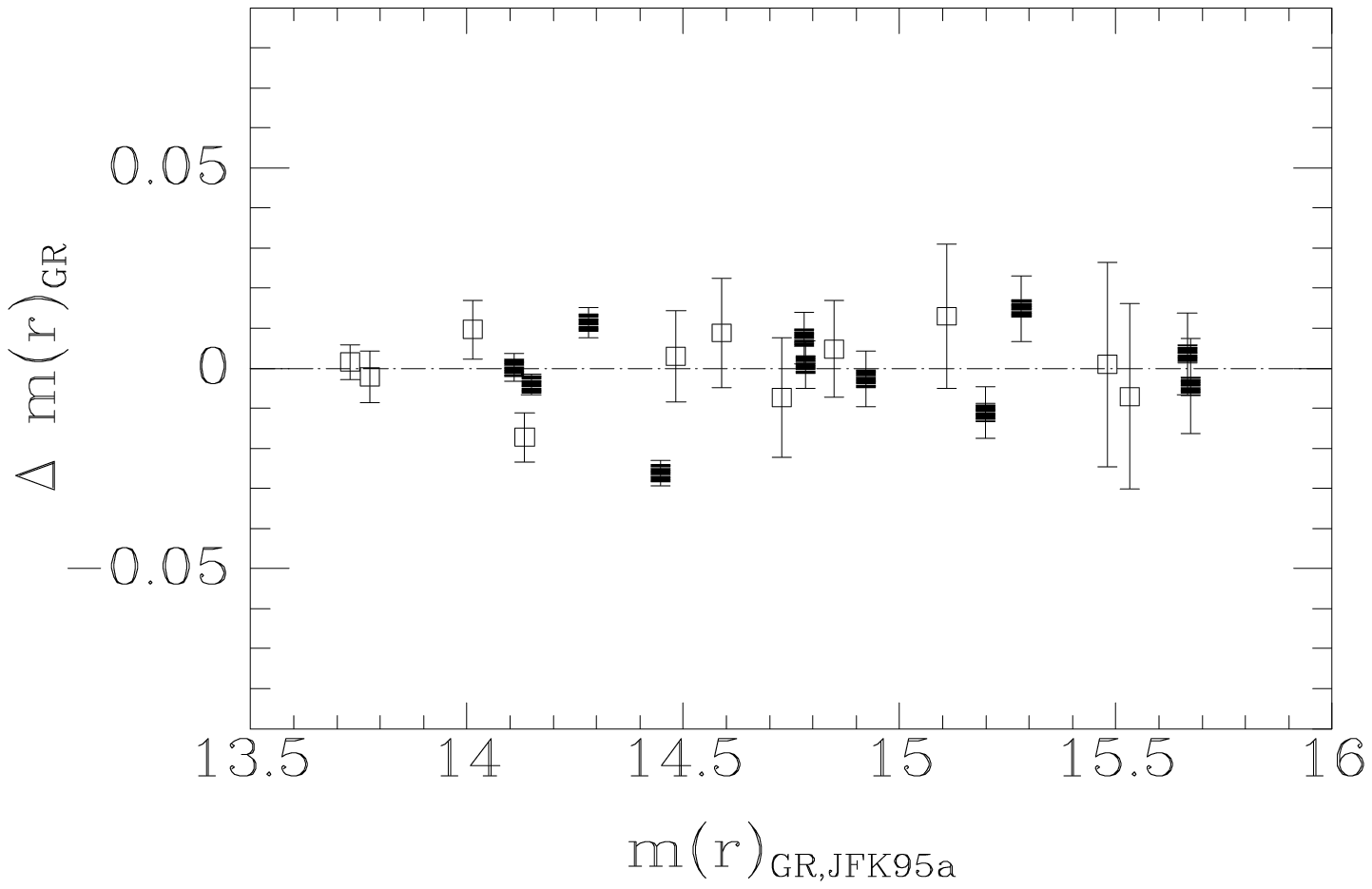}

\caption[]{
Comparison of aperture magnitudes in Gunn $r$ for galaxies in S639.
Filled boxes -- aperture radius $6''$; open boxes -- aperture radius 
$10''$.
The differences are calculated as ``JFK95a''-``this paper''. The
photometry presented in this paper have been offset to the mean
zero point defined by the JFK95a photometry.
The rms scatter of the comparison is $0\fm 01$.

\label{fig-apmags}
}
\end{figure}

The magnitudes were standard calibrated by means of offsets relative
to the photometry presented in JFK95a.
Figure \ref{fig-apmags} shows the comparison of aperture magnitudes
for the two sets of data, after the offset has been applied.
The rms scatter of the comparison is  $0\fm 01$.

Effective radii and surface brightnesses were derived by fitting
the growth curves of the galaxies with growth curves for $r^{1/4}$
profiles. The seeing was taken into account. We refer to 
JFK95a for further details.

We have compared the effective parameters derived from the new data
to those given in JFK95a. There are 11 galaxies in common. 
For differences derived as ``JFK95a''--``this paper'' we find 
mean offsets of
$\Delta \log \re = -0.006 \pm 0.019$ with an rms scatter of 0.063;
$\Delta \mue = -0.02 \pm 0.06$ with an rms scatter of 0.22;
$\Delta \mT = 0.01 \pm 0.03$ with an rms scatter of 0.10.
The combination $\log \re - 0.35 \mue$, which enters the FP
has also been compared. We find the difference to be
$0.001\pm 0.003 $ with an rms scatter of 0.011.
For the galaxies in common we use the average of the photometric 
parameters given in JFK95a and the new parameters derived here.

Table \ref{tab-phot} lists the average photometric parameters for
all the observed galaxies. The table also gives the spectroscopic
parameters used in the analysis. These data are from
JFK95b and J\o rgensen (1997).
Further, the $(B-r)$ colors derived from the measured $(g-r)$ colors
(JFK95a) are listed.

\begin{table*}
\begin{minipage}{11.0cm}
\caption[]{Photometric and spectroscopic parameters for S639 galaxies
\label{tab-phot} }

\begin{tabular}{lrrrrrrr}
Galaxy & $\log \re$ & $\mue$ & $(B-r)$ & $\log \sigma$ & 
$\Mgtwo$ & $\log \Fe$ & $\log \HbG$ \\ \hline
%
E264G23 & 1.24 & 20.63 & 1.14 & 2.342 & 0.291 & 0.429 & ...  \\
E264G24 & 0.97 & 18.92 & 1.06 & 2.349 & 0.273 & 0.502 & 0.392  \\
E264G26$^a$ & 1.18 & 20.74 & ... & 2.115 & 0.243 & 0.456 & ...  \\
E264G28$^a$ & 1.46 & 21.44 & ... & 2.191 & 0.242 & 0.490 & ...  \\
E264G300 & 1.18 & 20.01 & 1.15 & 2.449 & 0.315 & 0.462 & 0.304  \\
E264G301 & 0.66 & 18.91 & 1.11 & 2.265 & 0.280 & 0.497 & 0.400  \\
E264G302 & 0.60 & 18.83 & 1.17 & 2.276 & 0.262 & 0.424 & ...  \\
E264G31 & 1.28 & 20.35 & 1.11 & 2.392 & 0.283 & 0.509 & ...  \\
J06$^a$ & 0.99 & 20.41 & 1.12 & 2.041 & 0.201 & 0.403 & ...  \\
J09 & 0.59 & 20.43 & ... & 1.852 & 0.210 & 0.260 & ...  \\
J10 & 0.85 & 21.50 & ... & 1.856 & 0.192 & 0.463 & ...  \\
J13 & 0.84 & 19.08 & 1.13 & 2.298 & 0.263 & 0.471 & 0.352  \\
J14 & 0.89 & 19.76 & 1.09 & 2.135 & 0.244 & 0.442 & ...  \\
J15 & 0.49 & 18.73 & 1.09 & 2.105 & 0.207 & 0.422 & ...  \\
J16 & 0.61 & 19.60 & 1.11 & 2.094 & 0.254 & 0.455 & ...  \\
J18 & 0.39 & 19.91 & ... & 1.935 & 0.223 & 0.294 & ...  \\
J19$^a$ & 0.96 & 19.64 & 1.10 & ... & ... & ... & ...  \\
J20 & 0.52 & 20.38 & ... & 1.633 & 0.182 & 0.206 & ...  \\
J23 & 0.38 & 18.64 & 1.13 & 2.140 & 0.249 & 0.402 & ...  \\
J26 & 0.49 & 18.31 & ... & 2.496 & 0.303 & 0.484 & ...  \\
J31 & 0.81 & 20.98 & ... & 1.715 & ... & ... & ...  \\
J32 & 0.72 & 19.25 & ... & 2.187 & 0.267 & 0.510 & ...  \\
J101 & 0.73 & 20.76 & ... & 1.960 & 0.210 & 0.421 & ...  \\
J104 & 1.26 & 21.49 & ... & 2.018 & 0.137 & 0.207 & ...  \\
J109 & 1.15 & 21.00 & ... & 2.154 & 0.252 & 0.501 & ...  \\ \hline
\end{tabular}

Notes -- $^a$\, spiral galaxy, not included in the analysis.
$\re$ is in arcsec.
$\mue$ and $(B-r)$ have been corrected for the 
galactic extinction ($\AB$=0.75; $\Ar$=0.47), k-corrected,
and corrected for the cosmological dimming. The total correction 
for $\mue$ is $0\fm 586$ which has been subtracted from the standard 
calibrated magnitudes.
The total magnitudes can be calculated as 
$\mT = \mue - 5 \log \re - 2.5 \log 2\pi$.
The typical internal uncertainties on $\log \re$ and $\mue$ are 0.01
and 0.04, respectively.
The $(B-r)$ colors are derived from $(g-r)$ given by 
JFK95a using the relation $(B-r)=0.88(g-r)+0.73$.
The spectroscopic parameters are from JFK95b
and J\o rgensen (1997), see these references for uncertainties
on the parameters.

\end{minipage}
\end{table*}
 
\section{Str\"{o}mgren photometry for the stars}
 
Str\"{o}mgren photometry of S639 was obtained with
the Danish 1.5m telescope, La Silla, using the DFOSC.
We used a 2k$\times$2k thinned Loral chip (W11-4).
Due to problems with the stability of the flat fields 
the chip could not be used in the UV-flooded mode.
The stability problem of the flat fields was avoided by using the 
unflooded mode, but the resulting response in the $u$-band was $\la$10\%.
Furthermore, in unflooded mode the chip had a very low 
quantum efficiency (QE) near the edge of the frame.
As a consequence only the central 1450$\times$1450 pixels were used 
for the Str\"{o}mgren photometry. This area corresponds to 
$9\farcm 5 \times 9\farcm 5$ on the sky.

The exposure times were $2\times45$ minutes i $u$, 20 minutes in $v$,
10 minutes in $b$ and $y$, 12 minutes in $\beta_{\rm wide}$ and 
40 minutes in $\beta_{\rm narrow}$. 
Further, sequences of observations with shorter exposure times were 
obtained in order to tie-in the bright and faint stars in the field. 

The CCD images were reduced in a standard way, correcting for bias
and flat field corrected. 
Twi-light sky flat fields were used for all six filters.
The accuracy of the flat field correction is $\approx 1$\% in all 
filters except in $u$ where the effect of the low edge-QE is 
noticeable as an increase in the noise.

\subsection{Transformation to the standard system}

During the first seven nights of the observing period simultaneous
observations were performed using the Danish 0.5m telescope 
(the Str\"{o}mgren Automatic Telescope, SAT), La Silla. The SAT
was used for photoelectric observations of the $uvby$-$\beta$ primary
standard stars, some of the secondary standard stars used for
the CCD-photometry, and some bright stars in the CCD program
fields (though none in S639). The SAT also supplied nightly extinction
coefficients. For the rest of the period where no SAT observations
were available we used the mean extinction coefficients
for the first period.
One night was used for CCD observations of E-region standard stars 
(J\o nch-S\o rensen 1993). A total of 17 standard stars were observed. 
Transformations from
instrumental magnitudes to $V, ~(b-y), ~m_{1} \equiv (v-b)-(b-y),
 ~c_{1} \equiv (u-v)-(v-b)$ and $\beta$ were derived.
The rms scatter of the transformations were 
0.011, 0.010, 0.011, 0.057 and 0.013, respectively.
The transformations will be described in more detail in 
Milvang-Jensen et al.\ (1998).
The transformations are valid for stars from A-type to early G-type.
These stars have $2.58 \la \beta \la 2.90$, corresponding to
$0.43 \ga (b-y) \ga 0.0$ and luminosity classes V-III.

\subsection{The limiting magnitude and the accuracy of the magnitudes}

In the observed field 110 objects have complete $uvby$-$\beta$ data. 
The final sample used for the $E(b-y)$ analysis was limited to 
22 stars meeting the
restrictions implied by the applied calibrations (see Sect.\ B3).
The low sensitivity of the CCD at short wavelengths
limits the sample of stars with complete $uvby$-$\beta$ information
to stars brighter than $V=17\mag$. The limiting magnitude
in the other filters is $V\approx 20\mag$.
For stars brighter than $V=17\mag$ the internal uncertainties are
 $\sigma_{V}\! \le\! 0\fm 008$, $\sigma_{b-y}\! \le\! 0.012$,
$\sigma_{m_{1}}\!\le\! 0.018$, $\sigma_{c_{1}}\! \le\! 0.065$ and
$\sigma_{\beta}\! \le\! 0.008$. The poor UV-response
is manifested in the large uncertainty of the $c_{1}$ index.

\subsection{Determination of $E(b-y)$ and $\MV$}

Intrinsic colors can be estimated in the $uvby$-$\beta$ system for
individual stars of spectral types ranging from B- to G-type.
The $\beta$ index ($\equiv \beta_{narrow} - \beta_{wide})$ 
is not affected by reddening and is an effective temperature indicator 
for the stars in question. 
For definitions and properties of the $uvby$-$\beta$ system
see Crawford (1975).

In this program we concentrate on F- and early G-type stars since these
stars are frequent and the calibration of $uvby$-$\beta$ indices in 
terms of both intrinsic color and astrophysical parameters such as 
$\MV$, $\Teff$ and [Fe/H] are well-established.
We follow closely the procedure outlined in J\o nch-S\o rensen (1994ab).

We use the intrinsic color calibration of $(b-y)_{0}$ by Olsen (1988).
$(b-y)_{0}$ is calibrated using $\beta$ as indicator of the effective 
temperature with corrections of the luminosity and the metallicity via
terms involving $c_1$ and $m_1$, respectively (see Olsen 1988 for 
details).  The rather complex relation between the three indices 
means that the uncertainty on the derived $E(b-y)$ depends upon 
position in the ($\beta,m_1,c_1$) space. 
There is no tight relation between $V$ and and the uncertainty on
$E(b-y)$ as one might have expected from the relation between the
magnitude uncertainties and $V$.
However, there is an effect of the spectral type in the sense that
the uncertainty on $E(b-y)$ increases near the cool end of the sample. 
For $(b-y)_{0} \le 0.39$ the average uncertainty is 
$<\!\!\sigred\!\!>\,= 0.022$ while for $(b-y)_{0} \ge 0.4$  
the uncertainty is $<\!\!\sigred\!\!>\,= 0.070$.
The average uncertainty for our sample is
$<\!\!\sigred\!\!>\,= 0.039$.

The distances are estimated from $\MV$, which is derived using the
calibration by Crawford (1975). Again $\beta$ is the
main parameter with corrections of the luminosity from $c_1$.
The uncertainty of the derived $\MV$ is closely correlated with $V$,
increasing from $<\!\!\sigmv\!\!>\,= 0\fm 17$ for $V \le 14\mag$ to
$0\fm 93$ for $V\ge16\mag$.
This includes contributions from the uncertainties on $E(b-y)$ and $V$.
The uncertainties on $\MV$ correspond to (formal) relative distance 
uncertainties between $\approx$10\% and $\approx$45\%. 
Note, that the rms scatter of the calibration of photometric 
distances is $\approx$15\%, cf., Nissen (1994).


\begin{thebibliography}{}

\bibitem[\protect\citename{Abell et al.\ }1989]{abell:1989}
Abell G.\ O., Corwin H.\ G., Jr., Olowin, R.\ P., 1989, ApJS, 70, 1

\bibitem[\protect\citename{Baggley }1996]{baggley:1996}
Baggley G., 1996, PhD thesis, Oxford University

\bibitem[\protect\citename{Bender et al.\ }1993]{bender:1993}
Bender R., Burstein D., Faber S.\ M., 1993, ApJ, 411, 153

\bibitem[\protect\citename{Burstein \& Heiles }1982]{burstein:1982}
Burstein D., Heiles C., 1982, AJ, 87, 1165

\bibitem[\protect\citename{Burstein et al.\ }1987]{burstein:1987} 
Burstein D.,  Davies R.\ L., Dressler A., Faber S.\ M.,  
Stone R.\ P.\ S., Lynden-Bell D., Terle\-vich  R.\ J.,  Wegner G.,
1987,  ApJS,  64,  601
 
\bibitem[\protect\citename{Burstein et al.\ }1988]{burstein:1988}
Burstein D.,  Davies R.\ L.,  Dressler A., Faber S.\ M.,
Lynden-Bell D.,  Terle\-vich  R.\ J., Wegner G.,
1988, in Kron R.\ G., Renzini A., eds.,
Towards Understanding Galaxies at Large Redshifts,
Kluwer Academic Publishers, Dordrecht, p.\ 17
  
\bibitem[\protect\citename{Burstein et al.\ }1990]{burstein:1990} 
Burstein D., Faber S.\ M., Dressler A., 1990,  ApJ,  354,  18
  
\bibitem[\protect\citename{Crawford }1975]{crawford:1975}
Crawford D.\ L., 1975, AJ, 80, 955

\bibitem[\protect\citename{Davies et al.\ }1987]{davies:1987}
Davies R.\ L., Burstein D., Dressler A., Faber S.\ M., 
Lynden-Bell D., Terlevich R.\ J., Wegner G., 1987, ApJS, 64, 581
 
\bibitem[\protect\citename{de Carvalho \& Djorgovski }1992]{decarvalho:1992}
de Carvalho R.\ R., Djorgovski S., 1992, ApJ, 389, L49
 
\bibitem[\protect\citename{Djorgovski \& Davis }1987]{djorgovski:1987} 
Djorgovski S.,  Davis M., 1987,  ApJ,  313,  59
 
\bibitem[\protect\citename{Dressler }1987]{dressler:1987} 
Dressler A., 1987,  ApJ,  317,  1
  
\bibitem[\protect\citename{Dressler et al.\ }1987b]{dressler:1987} 
Dressler A.,  Lynden-Bell D.,  Burstein D.,  Davies R.\ L.,  
Faber S.\ M.,  Terlevich R.\ J.,  Wegner G.,  1987,  ApJ,  313,  42
 
\bibitem[\protect\citename{Faber et al.\ }1988]{faber:1988} 
Faber S.\ M.,  Burstein D., 1988,  
in Rubin V.\ C., Coyne G.\ V., eds.,
Large-Scale Motions in the Universe. A Vatican Study Week, 
Princeton University Press, Princeton, p.\ 115

\bibitem[\protect\citename{Faber et al.\ }1985]{faber:1985}
Faber S.\ M., Friel E.\ D., Burstein D., Gaskell C.\ M., 1985,
ApJS, 57, 711

\bibitem[\protect\citename{Franco }1989]{franco:1989}
Franco G.\ A.\ P., 1989, A\&A, 227, 499

\bibitem[\protect\citename{Franx et al.\ }1989]{franx:1989} 
Franx M., Illingworth G., Heckman T.,  1989,  AJ,  98,  538

\bibitem[\protect\citename{Gregg }1992]{gregg:1992}
Gregg M.\ D., 1992, ApJ, 384, 43

\bibitem[\protect\citename{Guzm\'{a}n et al.\ }1992]{guzman:1992}
Guzm\'{a}n R., Lucey J.\ R., Carter D., Terlevich R.\ J.,
1992, MNRAS, 257, 187

\bibitem[\protect\citename{Hilditch et al.\ }1983]{hilditch:1983}
Hilditch R.\ W., Hill G., Barnes J., 1983, MNRAS, 204, 41

\bibitem[\protect\citename{Hudson et al.\ }1997]{hudson:1997}
Hudson M.\ J., Lucey J.\ R., Smith R.\ J., Steel J., 1997,
MNRAS, 291, 488

\bibitem[\protect\citename{J{\o}nch-S{\o}rensen }1993]{HJS:1993}
J{\o}nch-S{\o}rensen H., 1993, A\&AS, 102, 637

\bibitem[\protect\citename{J{\o}nch-S{\o}rensen }1994a]{HJS:1994a}
J{\o}nch-S{\o}rensen H., 1994a, A\&AS, 108, 403

\bibitem[\protect\citename{J{\o}nch-S{\o}rensen }1994b]{HJS:1994b}
J{\o}nch-S{\o}rensen H., 1994b, A\&A, 292, 92

\bibitem[\protect\citename{J{\o}nch-S{\o}rensen \& Knude }1994]{HJS-JK:1994}
J{\o}nch-S{\o}rensen H., Knude J., 1994, A\&A, 288, 139

\bibitem[\protect\citename{J{\o}rgensen }1997]{IJ:1997}
J{\o}rgensen I., 1997, MNRAS, 288, 161

\bibitem[\protect\citename{J{\o}rgensen }1998]{IJ:1998}
J{\o}rgensen I., 1998, in preparation

\bibitem[\protect\citename{J{\o}rgensen et al.\ }1992]{IJ:1992} 
J{\o}rgensen I., Franx M.,  Kj{\ae}rgaard P., 1992,  A\&AS,
95, 489
  
\bibitem[\protect\citename{J{\o}rgensen et al.\ }1995a]{IJ:1995a} 
J{\o}rgensen I., Franx M., Kj{\ae}rgaard P., 1995a, MNRAS, 273, 1097
(JFK95a)
 
\bibitem[\protect\citename{J{\o}rgensen et al.\ }1995b]{IJ:1995b} 
J{\o}rgensen I., Franx M., Kj{\ae}rgaard P., 1995b, MNRAS, 276, 1341
(JFK95b)

\bibitem[\protect\citename{J{\o}rgensen et al.\ }1996]{IJ:1996} 
J{\o}rgensen I., Franx M., Kj{\ae}rgaard P., 1996, MNRAS, 280, 167
(JFK96)

\bibitem[\protect\citename{Knude }1979a]{knude:1979a}
Knude J., 1979a, A\&A, 71, 344

\bibitem[\protect\citename{Knude }1979b]{knude:1979b}
Knude J., 1979b, A\&A, 77, 198

\bibitem[\protect\citename{Knude }1981]{knude:1981}
Knude J., 1981, A\&A, 97, 380
 
\bibitem[\protect\citename{Lucey et al.\ }1991]{lucey:1991}
Lucey J.\ R., Guzm\'{a}n R., Carter D.,  Terlevich, R.\ J.,
1991, MNRAS, 253, 584

\bibitem[\protect\citename{Lynden-Bell et al.\ }1988]{lyndenbell:1988}
Lynden-Bell D., Faber S.\ M., Burstein D., Davies R.\ L.,
Dressler A., Terlevich R.\ J., Wegner G.,  1988,  ApJ,  326,  19
 
\bibitem[\protect\citename{Milvang-Jensen \& J\o rgensen }1998]{MJJ:1998}
Milvang-Jensen B., J\o rgensen I., 1998, in preparation

\bibitem[\protect\citename{Milvang-Jensen et al.\ }1998]{MJJJS:1998}
Milvang-Jensen B., J\o rgensen I., J\o nch-S\o rensen H., 1998, in preparation

\bibitem[\protect\citename{Nissen }1994]{nissen:1994}
Nissen P.\ E., 1994, Rev.\ Mex.\ Astron.\ Astrofis., 29, 129

\bibitem[\protect\citename{Olsen }1988]{olsen:1988}
Olsen E.\ H., 1988, A\&A, 189, 173

\bibitem[\protect\citename{Salpeter }1955]{salpeter:1955}
Salpeter E.\ E., 1955, ApJ, 121, 161

\bibitem[\protect\citename{Schlegel et al.\ }1998]{schlegel:1998}
Schlegel D.\ J., Finkbeiner D.\ P., Davis M., 1998, ApJ, in press
(astro-ph/9710327)

\bibitem[\protect\citename{Vazdekis et al.\ }1996]{vazdekis:1996}
Vazdekis A., Casuso E., Peletier R.\ F. Beckman, J.\ E.,
1996, ApJS, 106, 307

\bibitem[\protect\citename{Worthey et al.\ }1994]{worthey:1994}
Worthey G., Faber S.\ M., Gonz\'{a}les J.\ J., Burstein D.,
1994, ApJS, 94, 687

\bibitem[\protect\citename{Zabludoff et al.\ }1990]{zablu:1990}
Zabludoff A., Huchra J.\ P., Geller M.\ J., 1990, ApJS, 74, 1

\end{thebibliography}
\end{document}